\documentclass[12pt]{article}
\usepackage{epsfig}

\date{}

\newcommand{\beeq}{\begin{equation}}
\newcommand{\eneq}{\end{equation}}
\newcommand{\be}{\begin{eqnarray}}
\newcommand{\ee}{\end{eqnarray}}
\newcommand{\bpic}{\begin{picture}}
\newcommand{\epic}{\end{picture}}

\begin{document}

\bibliographystyle{unsrt}
\footskip 1.0cm

\thispagestyle{empty}

\vspace{1in}

\begin{center}{\Large \bf { Prompt photons at RHIC }}\\

\vspace{1in}
{\large  J. Jalilian-Marian$^1$, K. Orginos$^{1,2}$ and I. Sarcevic$^1$}\\

\vspace{.2in}
{\it $^1$Department of Physics, University of Arizona, Tucson, Arizona
85721\\
$^2$RIKEN-BNL Research Center,
Brookhaven National Laboratory,
Upton NY 11973-5000\\
}

\end{center}

\vspace*{25mm}

\begin{abstract}
\baselineskip=18pt
We calculate the inclusive cross section for prompt photon production 
in heavy-ion collisions at RHIC energies ($\sqrt{s}=130$ GeV and 
$\sqrt{s}=200$ GeV) in the central rapidity region including 
next-to-leading order, $O(\alpha_{em}\alpha_s^2)$, 
radiative corrections, initial state nuclear shadowing and
parton energy loss effects.  We show that there is a significant 
suppression of the nuclear cross section, up to $\sim 30\%$ at  
$\sqrt{s}=200$ GeV, due to shadowing and medium induced parton energy 
loss effects.  We find that the next-to-leading order contributions are 
large and have a strong $p_t$ dependence.  

\end{abstract}

\vspace*{5mm}

\newpage

\normalsize
There is much excitement about the recent long awaited 
first collisions at the 
Relativistic Heavy Ion Collider (RHIC) at Brookhaven National
Laboratory (BNL) due to the fact 
that the energy densities and
temperatures created are larger than those 
ever seen before in a laboratory environment.  It is
commonly believed that at such high energy density and temperature,
a new state of matter, the Quark Gluon Plasma (QGP), will be formed as
predicted by QCD. However, the signatures of the Quark Gluon Plasma
are non-trivial and hard to interpret. Electromagnetic probes such
as dileptons and photons are excellent tools to investigate the Quark
Gluon Plasma since once produced, they do not strongly interact with the
plasma and therefore carry information about all stages of the
heavy ion collisions \cite{qm}.

Photon production in heavy-ion collisions has been studied extensively 
\cite{phot,jk,hdsg,ss}.  
There are several sources of photon production in a heavy-ion 
collision: thermal photons which are emitted from a thermalized
plasma, photons from decays of hadrons such as pions and eta's, and
photons produced in early stages of the collision through hard processes.
Each of these processes 
dominates in a specific transverse momentum region.
Photon production from hard partonic processes is expected to be the 
dominant mechanism at high $p_t$ ($p_t > 4 $ GeV), 
while at lower values of $p_t$ ($p_t \sim  3-4$ GeV), 
it is an important background to thermal photons.  
Here we focus on photon
production from hard partonic processes. This is timely as 
PHENIX collaboration has just collected photon production data 
up to $6$ GeV \cite{mike}. Quantitative knowledge of the  
production cross section of hard photons 
is also extremely important in deciding whether or not
thermal photons can be used as a signal for the 
formation of the Quark Gluon Plasma at RHIC.  

In perturbative QCD the inclusive cross section for
prompt photon production in nuclear collisions is given by
\be
E \frac{d^3\sigma}{d^3p}(\sqrt s,p_T)
=\int
dx_{a}
\int dx_{b} \int dz \sum_{i,j}^{partons}F_{i/A}(x_{a},Q^{2})
F_{j/B}(x_{b},Q^{2}) D^A_{\gamma /k}(z,Q_f^2)
E_\gamma \frac{d^3\hat{\sigma}_{i+j\rightarrow \gamma +X}}
{d^3p_\gamma}
\nonumber
\ee
where $F_{i/A}(x,Q^{2})$'s are the parton distributions in a nucleus, 
$x_a$ and $x_b$ are the fractional momenta of incoming partons and 
$D^A_{\gamma /k}(z,Q_f^2)$ are the nuclear fragmentation functions.
The parton distributions in nuclei are modified compared to those
in hadrons beyond a simple scaling by $A$. This modification is known
as the nuclear shadowing, 
defined as 
$S(x,Q^2,A)\equiv \frac {F_{i}^A(x,Q^2)}{A F_{i}^N(x,Q^2)}$, 
and can be parameterized as \cite{bqv}: 
\begin{equation}
S(x,Q^2,A)
=\left\{\begin{array}{ll}
\alpha_3 -\alpha_4 x & x_0 <x\leq 0.6 \\
(\alpha_3 -\alpha_4 x_0)\frac{1+k_q \alpha_2 ({1/x}-1/x_{0})}
{1+k_q A^{\alpha_1}({1/x}-1/x_{0})}
&x\leq x_{0}\\
\end{array}
\right\}
\label{eq:shadow}
\end{equation}

The parameters $\alpha_i$ and $k$ are given in \cite{bqv}.  
This parameterization fits all the EMC, NMC and E665 data on nuclear 
shadowing \cite{ar}. All the scales that appear 
in the cross section, renormalization, 
factorization and fragmentation, were set to $p_t/2$
as in \cite{aur} where it was shown that these choices of scales lead to 
the best agreement of 
theoretical predictions with the data.  
In our calculation we use the MRS99 parameterization of parton distributions
in hadrons \cite{mrs99} (also used in \cite{aur}) modified by the 
shadowing ratio as defined in (\ref{eq:shadow}). We take $A$ to be 
$200$ throughout this work.  

The double differential parton-parton cross section for prompt 
photon production,  
$E \frac{d^3\hat\sigma}{d^3p}$,  
has been calculated to order $O(\alpha_{em}\alpha_s^2$) in Ref. \cite{aur}
to which we refer the reader for a complete list of diagrams contributing
to next-to-leading order (NLO) and 
also an extensive discussion of hadronic structure and 
fragmentation functions.
Parton distributions in hadrons \cite{mrs99} and the photon fragmentation 
function \cite{pfrag} have both been calculated up to NLO, while the 
nuclear shadowing effect and the photon 
fragmentation in the nuclear medium have only been obtained in the leading 
order. We calculate the 
$K$-factor, usually introduced as the contribution of the NLO relative to 
the LO, and 
show that it is not a constant but rather varies 
significantly 
with energy and transverse momentum of the photon. 
We will define our K-factor relative to the leading order plus the 
bremsstrahlung contributions since a subset of the bremsstrahlung 
contributions were included in prior work on prompt photons \cite{hdsg}.  

\begin{figure}[htp]
\centering
\setlength{\epsfxsize=10cm}
\centerline{\epsffile{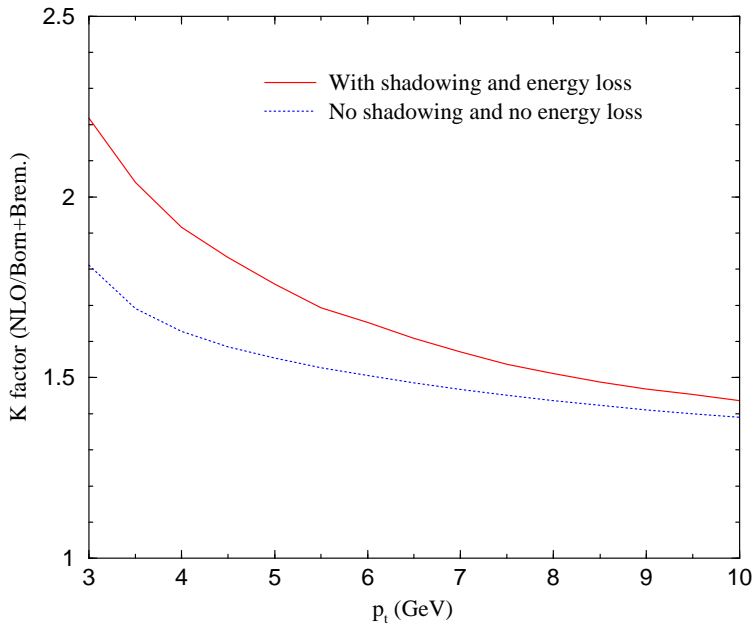}}
\caption{The K-factor for hadronic and nuclear collisions.} 
\label{fig:Kfactor}
\end{figure}

In Fig. \ref{fig:Kfactor} we show the $K$
factor, defined as $K\equiv$ NLO/(Born + Brems.) for prompt photon 
production in both hadronic and nuclear collisions.  The bremsstrahlung 
contribution in the denominator includes both leading and next-to-leading 
order bremsstrahlung diagrams.  

We note that $K$-factor in hadronic collisions varies from 
$1.4$ at $p_t=10$ GeV to $1.8$ at $p_t=3$ GeV.  However, the 
nuclear $K$-factor has even stronger $p_t$ dependence, ranging 
from $1.4$ at $p_t=10$ GeV to $2.2$ at $p_t=3$ GeV.  The 
difference comes from including the nuclear shadowing and 
parton energy loss in our calculation of the photon production 
in heavy-ion collisions. We also calculate the standard 
$K$-factor, defined as the ratio of the full NLO to the Born cross sections, 
i.e.  $K\equiv$ NLO/Born.  
We find this $K$-factor to be large, ranging from 
$6$ at $p_t=3$ GeV to $2$ at $p_t=10$ GeV in hadronic collisions 
and from $5.5$ at $p_t=3$ GeV to $1.9$ at $p_t=10$ GeV  in the nuclear case. 
Clearly, taking the $K$-factor to be a constant at all $p_t$ and at all 
energies is not justified and a full NLO calculation is important.  

We also need to consider the nuclear modification of the photon  
fragmentation functions \cite{pfrag} used in \cite{aur} for 
hadronic collisions. 
In order to determine the nuclear fragmentation function we follow 
a simple 
model of Wang, Huang and Sarcevic \cite{hsw} and modify 
the photon fragmentation function, $zD^0(z)$, which gives 
the probability for a parton to fragment into a photon, to include 
multiple scatterings of the fragmenting parton from the nuclear medium
before it fragments. The nuclear fragmentation function $zD(z)$ is given
in terms of the photon fragmentation function  $zD^0(z)$ by \cite{hsw} 
\be
zD_{A/a}(z,\Delta L,Q^2)=
\frac{1}{C^a_N}\sum_{n=0}^N P_a(n) \bigg[z^a_nD^0_{h/a}(z^a_n,Q^2)
+ \sum_{j=0}^{n} \bar{z}^j_aD^0_{h/g}(\bar{z}^j_a,Q^2)\bigg]
\label{eq:frag}
\ee
where $z^a_n=z/(1-(\sum_{i=0}^{n}\epsilon^a_i)/E_T)$, 
$\bar{z}_j^a=zE_T/\epsilon_j^a$ and $ P_a(n)$ is the probability that
a parton of flavor $a$ traveling a distance $\Delta L$ in the nuclear
medium will scatter $n$ times. It is given by

\be
 P_a(n) = \frac{(\Delta L/\lambda_a)^n}{n!} e^{-\Delta L/\lambda_a}, 
\ee
and 
$C^a_N=\sum_{n=0}^N P_a(n)$.  
For the 
energy loss per unit distance $\epsilon_a$, we take the energy dependent 
expression of Baier, Dokshitzer, Mueller, Peigne and Schiff \cite{eloss}, 
$\epsilon_a =\alpha_s \sqrt{{\mu^2 E \over \lambda_a}}$,  where
$E$ is the energy of the parton undergoing the multiple scatterings, 
$\lambda_a$ is the parton inelastic mean free path and 
$\mu^2$ represent a screening mass generated
by the plasma and serves as an infrared cut off. 
The first term in Eq. (2) 
corresponds to the 
fragmentation of the leading parton $a$ with reduced energy 
$E_T- \sum_{i=0}^{n}\epsilon^i_a$ after $n$ gluon emissions and the 
second term comes from the $j^{th}$ 
emitted gluon having energy $\epsilon^j_a$. 

\begin{figure}[hbp]
\centering
\setlength{\epsfxsize=10cm}
\centerline{\epsffile{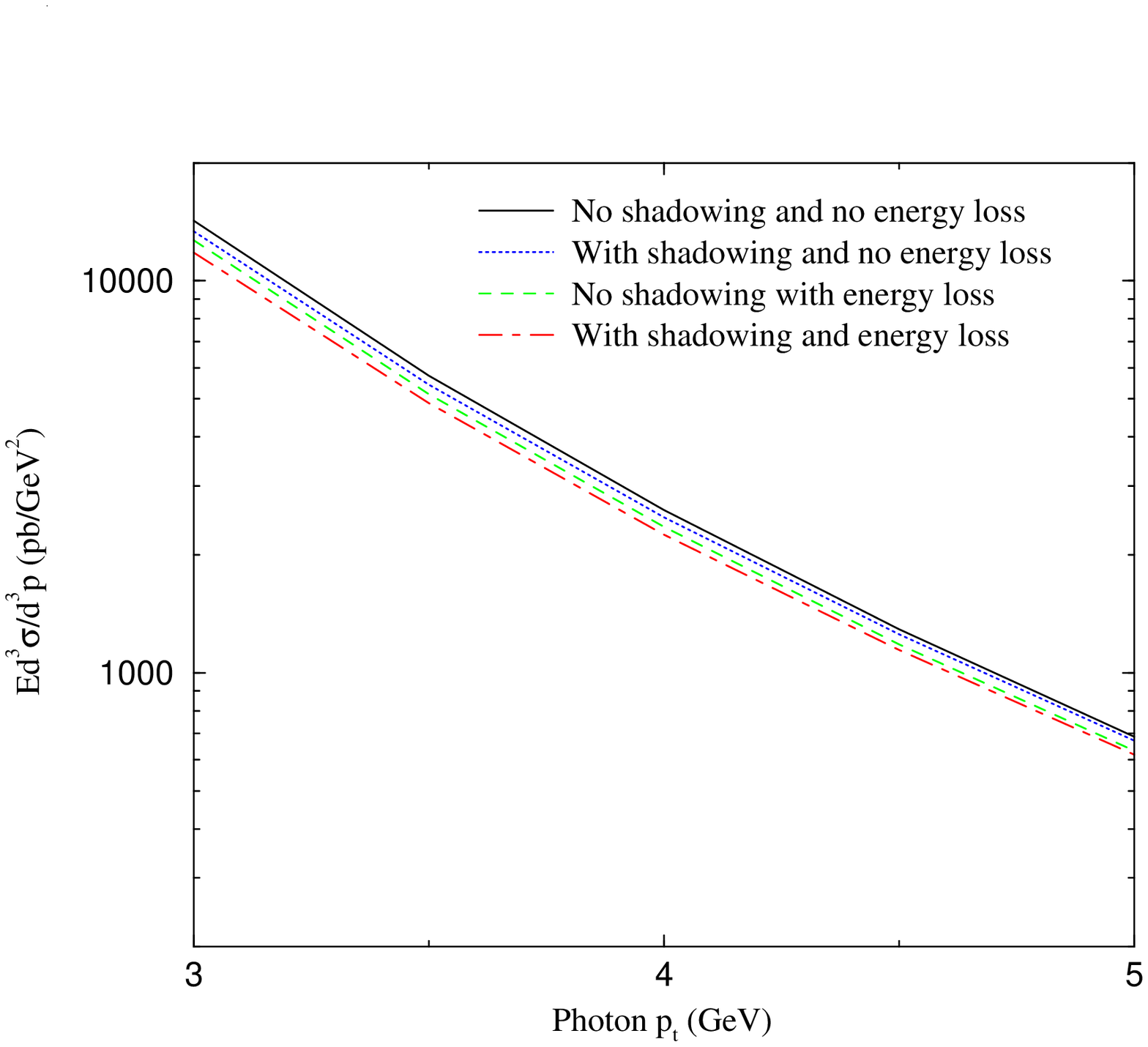}}
\caption{Prompt photon cross section at central rapidity at 
$\sqrt{s}=130$ GeV.} 
\label{fig:lrhic}
\end{figure}

In Fig. \ref{fig:lrhic} we show our results for the invariant 
photon cross section
as a function of photon $p_t$ for RHIC at $\sqrt{s}=130$ GeV. The effects
of nuclear shadowing and energy loss on the nuclear cross sections are
shown separately as well as combined. We take $\mu =1$ GeV and 
$\lambda_q = \lambda_g =1 fm$.  
We find that the 
nuclear shadowing and the parton energy loss effects result in 
about $18\%$ suppression of the cross section at $p_t=3$ GeV and
about $4\%$ at $p_t=10$ GeV. 

\begin{figure}[htp]
\centering
\setlength{\epsfxsize=10cm}
\centerline{\epsffile{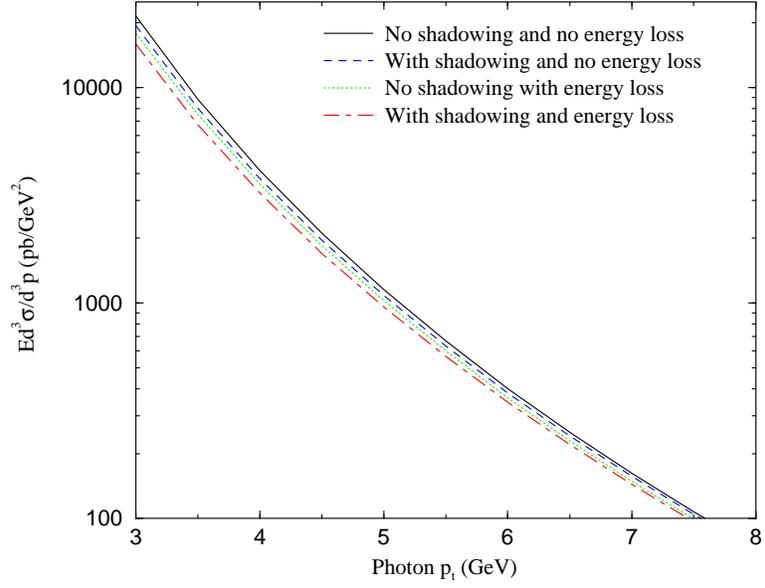}}
\caption{Prompt photon cross section at central rapidity 
at $\sqrt{s}=200$ GeV.} 
\label{fig:rhic}
\end{figure}

In Fig. \ref{fig:rhic} we show the same cross section as in 
Fig. \ref{fig:lrhic}. for RHIC
at the higher energy of $\sqrt{s}=200$ GeV, expected to be reached 
experimentally sometime this year.
As expected, the nuclear effects, shadowing and the parton energy loss 
become more
important at higher energies. At $p_t=3$ GeV, there is about 
$30\%$ 
suppression of the photon 
cross
section while at 
$p_t=10$ GeV it is 
a $7\%$ effect.  We expect that this 
effect can be observed experimentally at RHIC without ambiguity.

\begin{figure}[htp]
\centering
\setlength{\epsfxsize=10cm}
\centerline{\epsffile{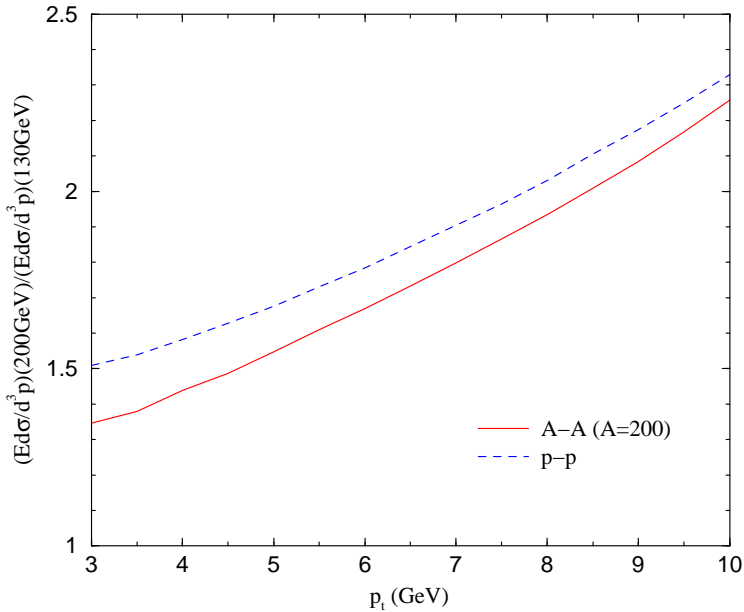}}
\caption{Ratio of hadronic and nuclear cross sections at $\sqrt{s}=130$ GeV
and $\sqrt{s}=200$~GeV.} 
\label{fig:rrhiclrhicvspt}
\end{figure}

In Fig. \ref{fig:rrhiclrhicvspt} we show the ratio of hadronic and 
nuclear cross
sections at different RHIC energies of $\sqrt{s}=130$ GeV and  
$\sqrt{s}=200$ GeV. By measuring this ratio one reduces the 
theoretical as well experimental uncertainties such as scale dependence
of cross sections and systematic errors. Clearly, 
nuclear effects are especially important
at lower $p_t$ where the measurements are expected in the near future.  

\begin{figure}[bhp]
\centering
\setlength{\epsfxsize=10cm}
\centerline{\epsffile{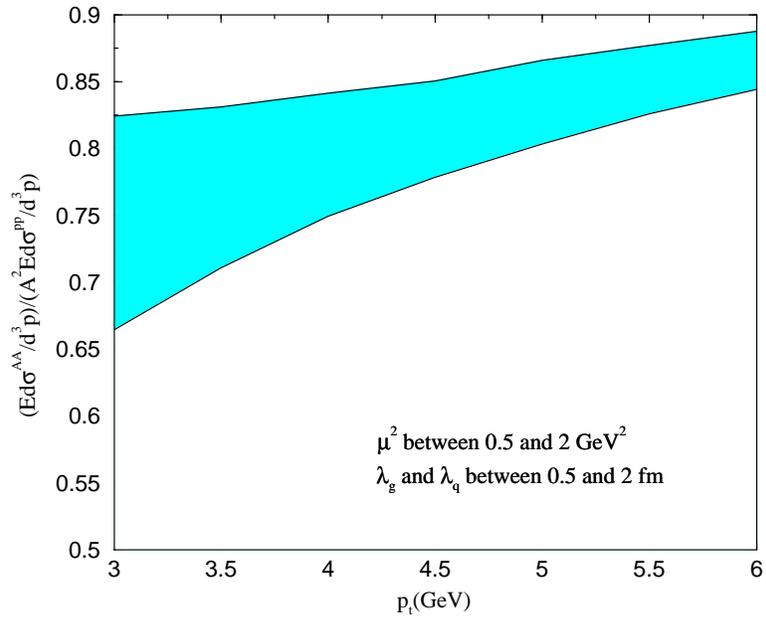}}
\caption{Uncertainty in the nuclear cross section due to variation
of the energy loss parameters at $\sqrt{s}=200$ GeV.} 
\label{fig:rsrhic11diffmulvspt}
\end{figure}

The dependence of our results on the choice of energy loss parameters 
$\mu^2$ and $\lambda_a$ is shown in Fig. \ref{fig:rsrhic11diffmulvspt} 
We have varied the 
values of $\mu^2$ and $\lambda$ as indicated in the figure. 
The upper limit corresponds to the case when $\mu =1$ GeV and 
$\lambda_q =\lambda_g =2 fm$ while the lower limit corresponds to
$\mu =1$ GeV and $\lambda_q =\lambda_g =0.5 fm$.
Variation of these parameters can result in as much uncertainty as $25\%$ for 
$\sqrt{s}=200$ GeV at the lowest $p_t$, but the overall effect is robust.

In summary, we have calculated the prompt photon production cross section
for heavy-ion collisions at RHIC including the 
next-to-leading order corrections, $O(\alpha_{em} \alpha_s^2)$, 
nuclear shadowing effect and the 
final state parton energy losses.  
We have found that higher-order corrections, even relative to leading 
order plus 
the bremsstrahlung, are large and depend on $p_t$ of the photon.  
In case of heavy-ion collisions, the $K$-factor has even stronger 
$p_t$ dependence, 
and thus should not be taken as a constant.  Taking $K$-factor as a 
constant gives 
a flatter cross section than the full next-to-leading order calculation.  
Measurement of prompt photons at high $p_t$ ($p_t > 4$ GeV) should serve 
as the 
normalization, as these photons dominate over thermal photons or photons from 
hadronic decays 
in this region.  However, 
at lower $p_t$ ($p_t \sim  3-4$ GeV) these photons constitute a 
significant background
to thermal photons and photons from hadronic decays.  
We have also  shown that nuclear effects are significant
and can be as much as $30\%$ at $\sqrt{s}=200$ GeV. Therefore, the PHENIX
collaboration should be able to, for the first time ever, see the parton energy
loss effects due to nuclear media. Extension of our work to p-A collisions 
at RHIC and heavy-ion collisions at higher energies such as the LHC 
will be reported elsewhere \cite{jos}.   
Calculation of prompt photon cross sections
in heavy-ion collisions at LHC energies is especially challenging since 
one
explores the small-$x$ region of phase space where gluons are dominant.  
Currently there is no experimental data on
nuclear structure functions in the small-$x$ region of relevance to LHC.
High parton density and higher twist effects such as those discussed
in \cite{smallx} will become important and need to be included.
Furthermore, the photon fragmentation functions are also not known at 
very small values of $z$ which will be explored by LHC. These issues are 
presently under investigation and results for prompt photons in heavy-ion 
collisions at LHC energies will be presented elsewhere \cite{jos}.

\leftline{\bf Acknowledgments} 

We would like to thank P. Aurenche and M. Werlen for providing us with 
the fortran routines for calculating double differential
distributions for photon production in hadronic collisions and for 
many useful discussions.  
J. J-M. would like to thank S. Jeon and X-N. Wang for discussions
on various topics related to this work. 
This work was supported in part through U.S. Department of Energy Grants Nos. 
DE-FG03-93ER40792 and DE-FG02-95ER40906.  
     
\leftline{\bf References}

\renewenvironment{thebibliography}[1]
        {\begin{list}{[$\,$\arabic{enumi}$\,$]}  
        {\usecounter{enumi}\setlength{\parsep}{0pt}
         \setlength{\itemsep}{0pt}  \renewcommand{\baselinestretch}{1.2}
         \settowidth
        {\labelwidth}{#1 ~ ~}\sloppy}}{\end{list}}

\end{document}